\documentclass[10pt,aps,twocolumn,prd,showpacs,nofootinbib,noshowkeys,floatfix,preprintnumbers]{revtex4-1}
\usepackage{graphics,graphicx}
\usepackage{amsmath, amssymb}
\usepackage{multirow}
\usepackage{color}
\usepackage{verbatim}
\usepackage{hyperref}
\usepackage[normalem]{ulem}  
\usepackage{ulem}
\usepackage{color}
\usepackage{cancel}
\usepackage{mathtools}
\usepackage{epstopdf}

\renewcommand\sout{\bgroup \color{blue}\ULdepth=-.5ex \ULset}

\renewcommand\sout{\bgroup \color{blue} \ULdepth=-.5ex \ULset}
\def\slashchar#1{\setbox0=\hbox{$#1$}  
\dimen0=\wd0     
\setbox1=\hbox{/} \dimen1=\wd1  
\ifdim\dimen0>\dimen1   
\rlap{\hbox to \dimen0{\hfil/\hfil}} 
#1     
\else     
\rlap{\hbox to \dimen1{\hfil$#1$\hfil}} 
/      
\fi}

\newcommand{\dd}{\mathrm{d}}
\newcommand{\pp}{\partial}

\begin{document}

\title{Interplay between chiral dynamics and repulsive interactions}
\date{\today}
\author{Micha\l{} Marczenko}
\email{michal.marczenko@uwr.edu.pl}
\affiliation{Institute of Theoretical Physics, University of Wroc\l{}aw, plac Maksa Borna 9, PL-50204 Wroc\l{}aw, Poland}
\author{Krzysztof Redlich}
\affiliation{Institute of Theoretical Physics, University of Wroc\l{}aw, plac Maksa Borna 9, PL-50204 Wroc\l{}aw, Poland}
\author{Chihiro Sasaki}
\affiliation{Institute of Theoretical Physics, University of Wroc\l{}aw, plac Maksa Borna 9, PL-50204 Wroc\l{}aw, Poland}

\begin{abstract}
  We investigate fluctuations of the net-baryon number density in hot hadronic matter. We discuss the interplay between chiral dynamics and repulsive interactions and their influence on the properties of these fluctuations near the chiral crossover. The chiral dynamics is modeled by the parity doublet Lagrangian that incorporates the attractive and repulsive interactions mediated via the exchange of scalar and vector mesons, respectively. The mean-field approximation is employed to account for chiral criticality. We focus on the properties and systematics of the cumulants of the net-baryon number density up to the sixth order. It is shown that the higher-order cumulants exhibit a substantial suppression in the vicinity of the chiral phase transition due to the presence of repulsive interactions. We find, however, that the characteristic properties of cumulants near the chiral crossover observed in lattice QCD results are entirely linked to the critical chiral dynamics and, in general, cannot be reproduced in phenomenological models, which account only for effective repulsive interactions via excluded-volume corrections or van-der-Waals type interactions. Consequently, a description of the higher-order cumulants of the net-baryon density in the chiral crossover requires a self-consistent treatment of the chiral in-medium effects and repulsive interactions.
\end{abstract}
\keywords{}
\pacs{}
\maketitle 

\section{Introduction}
\label{sec:introduction}

Establishing the thermodynamic properties of strongly interacting matter, described by quantum chromodynamics (QCD), is one of the key directions in modern high-energy physics. At vanishing density, the first-principles calculations of lattice QCD (LQCD) provide a reliable description of the equation of state (EoS) and fluctuations of conserved charges~\cite{Bazavov:2014pvz,Borsanyi:2018grb,Bazavov:2017dus,Bazavov:2020bjn,Bazavov:2020bjn}. There, the EoS exhibits a smooth crossover from confined hadronic matter to a deconfined quark-gluon plasma, which is linked to the color deconfinement and the restoration of chiral symmetry~\cite{Aoki:2006we, Bazavov:2018mes}. However, the nature of the EoS at finite density is still not resolved by LQCD, owing to the sign problem, and remains an open question.

The LQCD results~\citep{Aarts:2015mma, Aarts:2017rrl, Aarts:2018glk} exhibit a clear manifestation of the parity doubling structure for the low-lying baryons around the chiral crossover. The masses of the positive-parity ground states are found to be rather temperature-independent, while the masses of negative-parity states drop substantially when approaching the chiral crossover temperature $T_c$. The parity doublet states become almost degenerate with a finite mass in the vicinity of the chiral crossover. Even though these LQCD results are still not obtained in the physical limit, the observed behavior of parity partners is likely an imprint of the chiral symmetry restoration in the baryonic sector of QCD. Such properties of the chiral partners can be described in the framework of the parity doublet model~\citep{Detar:1988kn, Jido:1999hd, Jido:2001nt}. The model has been applied to hot and dense hadronic matter, neutron stars, as well as the vacuum phenomenology of QCD~\citep{Dexheimer:2007tn,Gallas:2009qp,Paeng:2011hy, Sasaki:2011ff, Gallas:2011qp, Zschiesche:2006zj, Benic:2015pia, Marczenko:2017huu, Marczenko:2018jui, Marczenko:2019trv, Marczenko:2020wlc, Marczenko:2020jma, Motornenko:2019arp, Mukherjee:2017jzi, Mukherjee:2016nhb, Dexheimer:2012eu, Steinheimer:2011ea, Weyrich:2015hha, Sasaki:2010bp, Yamazaki:2018stk,Yamazaki:2019tuo, Ishikawa:2018yey, Steinheimer:2010ib, Giacosa:2011qd, Motornenko:2018hjw,Motohiro:2015taa,Minamikawa:2020jfj}.

It is already confirmed,  that at small net-baryon number density,  the thermodynamics of the confined phase of QCD is well-described by the hadron resonance gas (HRG) model~\cite{BraunMunzinger:2003zd, Andronic:2017pug}. Since the fundamental quarks and gluons are confined, it is to be expected that at low temperatures the QCD partition function is dominated by the contribution of hadrons. The HRG model describes well the LQCD data below the crossover transition to a quark-gluon plasma, as well as the hadron yields in heavy-ion collisions~\cite{Andronic:2017pug}. Different extensions of the HRG model have been proposed to quantify the LQCD EoS and various fluctuation observable up to near chiral-crossover. They account for consistent implementation of hadronic interactions within the S-matrix approach~\cite{Venugopalan:1992hy, Broniowski:2015oha, Friman:2015zua, Huovinen:2016xxq, Lo:2017lym}, a more complete implementation or a continuously growing exponential mass spectrum and/or possible repulsive interactions among constituents~\cite{Majumder:2010ik, Andronic:2012ut,Albright:2014gva, Vovchenko:2014pka,Lo:2015cca, ManLo:2016pgd,Andronic:2020iyg}. Recently, an interesting suggestion was made that deviations of the LQCD data on higher-order fluctuations of net-baryon number density from the HRG  baseline in the near vicinity of the chiral transition can be attributed to repulsive interactions among constituent hadrons~\cite{Vovchenko:2016rkn}.

Fluctuations of conserved charges are known to be auspicious observable for the search of the chiral-critical behavior at the QCD phase boundary~\cite{Stephanov:1999zu, Asakawa:2000wh, Hatta:2003wn}, and chemical freeze-out of produced hadrons in heavy-ion collisions~\cite{Bazavov:2012vg, Borsanyi:2014ewa, Karsch:2010ck, Braun-Munzinger:2014lba, Vovchenko:2020tsr, Braun-Munzinger:2020jbk}. In particular, fluctuations have been proposed to probe the QCD critical point in the beam energy scan (BES) programs at the Relativistic Heavy Ion Collider (RHIC) at BNL and the Super Proton Synchrotron (SPS) at CERN, as well as the remnants of the $O(4)$ criticality at vanishing and finite baryon densities~\cite{Friman:2011pf, Karsch:2019mbv, Braun-Munzinger:2020jbk, Braun-Munzinger:2016yjz}. 

In this work, we analyze the properties and systematics of the fluctuations of conserved charges in the context of the parity doublet model, which incorporates the chiral symmetry restoration and repulsive interactions via the exchange of the scalar and vector mesons, respectively. To account for critical behavior, the mean-field approximation is employed, which contains basic features of the $O(4)$ criticality, albeit with different critical exponents. We study the behavior of the second- and higher-order cumulants of the net-baryon number density up to the sixth order, as well as the bulk equation of state. It is systematically examined to what extent the thermal behaviors are dominated by the chiral criticality and repulsive interactions.

This paper is organized as follows. In Sec.~\ref{sec:pd_model}, we introduce the parity doublet model. In Sec.~\ref{sec:cumulants}, we discuss the structure of the higher-order cumulants of the net-baryon number density. In Sec.~\ref{sec:results}, we present results on the equation of state and the higher-order cumulants of the net-baryon number density. Finally, Sec.~\ref{sec:summary} is devoted to summary and conclusions.

\section{Parity doublet model}
\label{sec:pd_model}

	In the conventional Gell-Mann--Levy model of mesons and nucleons~\cite{GellMann:1960np}, the nucleon mass is entirely generated by the non-vanishing expectation value of the sigma field. Thus, the nucleon inevitably becomes massless when the chiral symmetry gets restored. This is led by the particular chirality-assignment to the nucleon parity doublers, where the nucleons are assumed to be transformed in the same way as the quarks are under chiral rotations.

	More general allocation of the left- and right-handed chiralities to the nucleons, the mirror assignment, was proposed in~\cite{Detar:1988kn}. This allows an explicit mass-term for the nucleons, and consequently, the nucleons stay massive at the chiral restoration point. For more details, see Refs.~\cite{Detar:1988kn,Jido:1999hd,Jido:2001nt}.
	
	\begin{table*}[t!]\begin{center}\begin{tabular}{|c|c|c|c|c|c|c|c|c|c|c|c|}
        \hline
        $m_0~$[MeV] & $m_+~$[MeV] & $m_-~$[MeV] & $m_\pi~$[MeV] & $f_\pi~$[MeV] & $m_\omega~$[MeV] & $m_\rho~$[MeV] & $\lambda_4$ & $\lambda_6f_\pi^2$ & $g_\omega$ & $g_1$ & $g_2$ \\ \hline\hline
        850 & 939   & 1500  & 140     & 93      & 783        & 775 & 13.15          & 4.67           & 4.64       & 12.42 & 6.39 \\ \hline
        \end{tabular}\end{center}
        \caption{Physical vacuum inputs and the parity doublet model parameters used in this work. See Sec.~\ref{sec:pd_model} for details.}
        \label{tab:vacuum_params}
    \end{table*}

	In the mirror assignment, under \mbox{$SU(2)_L \times SU(2)_R$} rotation, two chiral fields $\psi_1$ and $\psi_2$ are transformed as follows:
	\begin{equation}\label{eq:mirror_assignment}
	\begin{split}
		\psi_{1L} \rightarrow L\psi_{1L}, \;\;\;\; \psi_{1R} \rightarrow R\psi_{1R}\textrm, \\
		\psi_{2L} \rightarrow R\psi_{2L}, \;\;\;\; \psi_{2R} \rightarrow L\psi_{2R}\textrm,
	\end{split}
	\end{equation}
	where $\psi_i = \psi_{iL} + \psi_{iR}$, $L \in SU(2)_L$ and $R \in SU(2)_R$. The nucleon part of the Lagrangian in the mirror model reads
	\begin{equation}\label{eq:doublet_lagrangian}
	\begin{split}
		\mathcal{L}_N &= i\bar\psi_1\slashchar\partial\psi_1 + i\bar\psi_2\slashchar\partial\psi_2 + m_0\left(  \bar\psi_1\gamma_5\psi_2 - \bar\psi_2\gamma_5\psi_1 \right) \\
		&+ g_1\bar\psi_1 \left( \sigma + i\gamma_5 \boldsymbol\tau \cdot \boldsymbol\pi \right)\psi_1 + g_2\bar\psi_2 \left( \sigma - i\gamma_5 \boldsymbol\tau \cdot \boldsymbol\pi \right)\psi_2 \\
		&-g_\omega\bar\psi_1\slashchar\omega\psi_1 - g_\omega\bar\psi_2\slashchar\omega\psi_2 \textrm,
	\end{split}
	\end{equation}
	where $g_1$, $g_2$, and $g_\omega$ are the baryon-to-meson coupling constants and $m_0$ is a mass parameter.

	The mesonic part of the Lagrangian reads
	\begin{equation}
	\begin{split}
		\mathcal{L}_M = \frac{1}{2} \left( \partial_\mu \sigma\right)^2 + \frac{1}{2} \left(\partial_\mu \boldsymbol\pi \right)^2 - \frac{1}{4} \left( \omega_{\mu\nu}\right)^2-V_\sigma - V_\omega \textrm,
	\end{split}
	\end{equation}
	where $\omega_{\mu\nu} = \partial_\mu\omega_\nu - \partial_\nu\omega_\mu$ is the field-strength tensor of the vector field, and the potentials read
	\begin{subequations}\label{eq:potentials_parity_doublet}
	\begin{align}
		V_\sigma &= -\frac{\lambda_2}{2}\Sigma + \frac{\lambda_4}{4}\Sigma^2 - \frac{\lambda_6}{6}\Sigma^3- \epsilon\sigma \textrm,\label{eq:potentials_sigma}\\
    	V_\omega &= -\frac{m_\omega^2 }{2}\omega_\mu\omega^\mu\textrm.
	\end{align}
	\end{subequations}
	where $\Sigma = \sigma^2 + \boldsymbol\pi^2$, $\lambda_2 = \lambda_4f_\pi^2 - \lambda_6f_\pi^4 - m_\pi^2$, and $\epsilon = m_\pi^2 f_\pi$. $m_\pi$ and $m_\omega$ are the $\pi$ and $\omega$ meson masses, respectively, and $f_\pi$ is the pion decay constant. Note that the chiral symmetry is explicitly broken by the linear term in $\sigma$ in Eq.~\eqref{eq:potentials_sigma}.

	The full Lagrangian of the parity doublet model is then
	\begin{equation}
		\mathcal L = \mathcal L_N + \mathcal L_M\textrm.
	\end{equation}

	In the diagonal basis, the masses of the chiral partners, $N_\pm$, are given by
	\begin{equation}\label{eq:doublet_masses}
		m_\pm = \frac{1}{2} \left( \sqrt{\left(g_1+g_2\right)^2\sigma^2+4m_0^2} \mp \left(g_1 - g_2\right)\sigma \right) \textrm.
	\end{equation}
	From Eq.~(\ref{eq:doublet_masses}), it is clear that, in contrast to the naive assignment under chiral symmetry, the chiral symmetry breaking generates only the splitting between the two masses. When the symmetry is restored, the masses become degenerate, $m_\pm(\sigma=0) = m_0$.

	To investigate the properties of strongly-interacting matter, we adopt a mean-field approximation~\cite{Serot:1984ey}. Rotational invariance requires that the spatial component of the $\omega_\mu$ field vanishes, namely $\langle \boldsymbol \omega \rangle = 0$\footnote{Since $\omega_0$ is the only non-zero component in the mean-field approximation, we simply denote it by $\omega_0 \equiv\omega$.}. Parity conservation on the other hand dictates $\langle \boldsymbol \pi \rangle = 0$.
	The mean-field thermodynamic potential of the parity doublet model reads
	\begin{equation}\label{eq:thermo_potential}
		\Omega = \sum_{x=\pm}\Omega_x + V_\sigma + V_\omega \textrm,
	\end{equation}
	with
	\begin{equation}\label{eq:kinetic_thermo}
		\Omega_x = \gamma_x \int\frac{\dd^3 p}{(2\pi)^3}\; T \left[ \ln\left(1 - f_x\right) + \ln\left(1 - \bar f_x\right) \right]\textrm,
	\end{equation}
	where $\gamma_\pm = 2\times 2$ denotes the spin-isospin degeneracy factor for both parity partners, and $f_x$  $(\bar f_x)$ is the particle (antiparticle) Fermi-Dirac distribution function,
	\begin{equation}\label{eq:fermi_dist_nucleon}
	\begin{split}
		f_x = \frac{1}{1+ e^{\beta\left(E_x - \mu^\ast\right)}} \textrm,\\
		\bar f_x = \frac{1}{1+ e^{\beta\left(E_x + \mu^\ast\right)}}\textrm, \\
	\end{split}
	\end{equation}
	with $\beta$ being the inverse temperature, the dispersion relation $E_x = \sqrt{\boldsymbol p^2 + m_x^2}$ and the effective chemical potential $\mu^\ast = \mu_B - g_\omega \omega$.

	In-medium profiles of the mean fields are obtained by extremizing the thermodynamic potential in Eq.~\eqref{eq:thermo_potential}, leading to the following gap equations:
	\begin{equation}
	\begin{split}\label{eq:gap_eqs}
		0=\frac{\partial \Omega}{\partial \sigma} &= \frac{\partial V_\sigma}{\partial \sigma} + s_+ \frac{\partial m_+}{\partial \sigma} + s_- \frac{\partial m_-}{\partial \sigma} \textrm,\\
		0=\frac{\partial \Omega}{\partial \omega} &= \frac{\partial V_\omega}{\partial \omega} + g_\omega \left(n_+ + n_-\right) \textrm,
	\end{split}
	\end{equation}
	where the scalar and vector densities are
	\begin{equation}
		s_\pm = \gamma_\pm \int \frac{\mathrm{d}^3 p}{\left(2\pi\right)^3} \frac{m_\pm}{E_\pm}\left(f_\pm + \bar f_\pm\right)
	\end{equation}
	and
	\begin{equation}
		n_\pm = \gamma_\pm \int \frac{\mathrm{d}^3 p}{\left(2\pi\right)^3}\left(f_\pm - \bar f_\pm\right) \textrm,
	\end{equation}
	respectively.

	In the grand canonical ensemble, the thermodynamic pressure reads
	\begin{equation}\label{eq:pressure}
		P= -\Omega + \Omega_0\textrm,
	\end{equation}
	where $\Omega_0$ is the value of the thermodynamic potential in the vacuum, and the net-baryon number density can be calculated as follows:
	\begin{equation}\label{eq:nb}
		n_B = \frac{\pp P(T, \mu_B)}{\pp \mu_B} \textrm.
	\end{equation}
	
	The positive-parity state, $N_+$, corresponds to the nucleon $N(938)$. Its negative parity partner is identified with $N(1535)$. Their vacuum masses are shown in Table~\ref{tab:vacuum_params}. The value of the parameter $m_0$ has to be chosen so that a chiral crossover transition is featured at finite temperature and vanishing chemical potential. The model predicts the chiral transition to be a crossover for $m_0\gtrsim 700~$MeV. Following the previous studies of the \mbox{parity-doublet-based} models~\citep{Dexheimer:2007tn,Gallas:2009qp,Paeng:2011hy, Sasaki:2011ff, Gallas:2011qp, Zschiesche:2006zj, Benic:2015pia, Marczenko:2017huu, Marczenko:2018jui, Marczenko:2019trv, Marczenko:2020wlc, Marczenko:2020jma, Motornenko:2019arp, Mukherjee:2017jzi, Mukherjee:2016nhb, Dexheimer:2012eu, Steinheimer:2011ea, Weyrich:2015hha, Sasaki:2010bp, Yamazaki:2018stk,Yamazaki:2019tuo, Ishikawa:2018yey, Steinheimer:2010ib, Giacosa:2011qd, Motornenko:2018hjw,Motohiro:2015taa,Minamikawa:2020jfj}, as well as recent lattice QCD results~\citep{Aarts:2017rrl, Aarts:2018glk,Aarts:2015mma}, we choose a rather large value, $m_0=850$~MeV. We note that, the results presented in Sec.~\ref{sec:results} qualitatively do not depend on the choice of $m_0$, as long as chiral crossover transition is featured. The parameters $g_1$ and $g_2$ are determined by the aforementioned vacuum nucleon masses and the chirally invariant mass $m_0$ via Eq.~\eqref{eq:doublet_masses}. The parameters $g_\omega$, $\lambda_4$ and $\lambda_6$ are fixed by the properties of the nuclear ground state at zero temperature, i.e., the saturation density, binding energy and compressibility parameter at $\mu_B=923~$MeV. The constrains are as follows:
    \begin{subequations}
    \begin{align}
        n_B &= 0.16~\textrm{fm}^{-3}\textrm,\\
        E/A - m_+ &= -16~\textrm{MeV}\textrm,\\
        K = 9n^2_B\frac{\pp^2\left( E/A \right)}{\pp n_B^2} &= 240~\textrm{MeV}\textrm.\label{eq:compres}
    \end{align}
    \end{subequations}
    We note that the six-point scalar interaction term in Eq.~\eqref{eq:potentials_sigma} is essential in order to reproduce the empirical value of the compressibility in Eq.~\eqref{eq:compres}~\citep{Motohiro:2015taa}. The parameters used in this paper are tabulated in Table~\ref{tab:vacuum_params}. For this set of parameters, we obtain the critical temperature of the chiral crossover transition at vanishing chemical potential, $T_c=208.7~$MeV.

	In the following, we will also compare our results with the hadron resonance gas (HRG)~\cite{Karsch:2003vd,Karsch:2003zq,Karsch:2013naa,Andronic:2012ut,Albright:2014gva,Albright:2015uua} model formulation of the thermodynamics of the confined phase of QCD. The thermodynamic potential of the HRG model is given as a sum of uncorrelated ideal-gas particles:
	\begin{equation}\label{eq:hrg_thermo}
		\Omega^{\rm HRG} = \sum_{x=\pm} \Omega_x \textrm,
	\end{equation}
	with $\Omega_x$ given by Eq.~\eqref{eq:kinetic_thermo}, where the masses of $N_\pm$ are taken to be the vacuum masses (see Table~\ref{tab:vacuum_params}) and $\mu^\ast = \mu_B$. We will also consider a modification of the HRG model, $\sigma$HRG, where the thermodynamic potential is given as in Eq.~\eqref{eq:hrg_thermo}, but the vacuum masses of $N^\pm$ are substituted by the in-medium masses obtained by solving the parity doublet model. In both models, the pressure and net-baryon density are obtained through Eqs.~\eqref{eq:pressure}~and~\eqref{eq:nb}, respectively.

	In the next section, we discuss the general structure of the higher-order cumulants of the net-baryon number.
	
	\begin{figure*}[t!]
		\centering
		\includegraphics[width=0.49\linewidth]{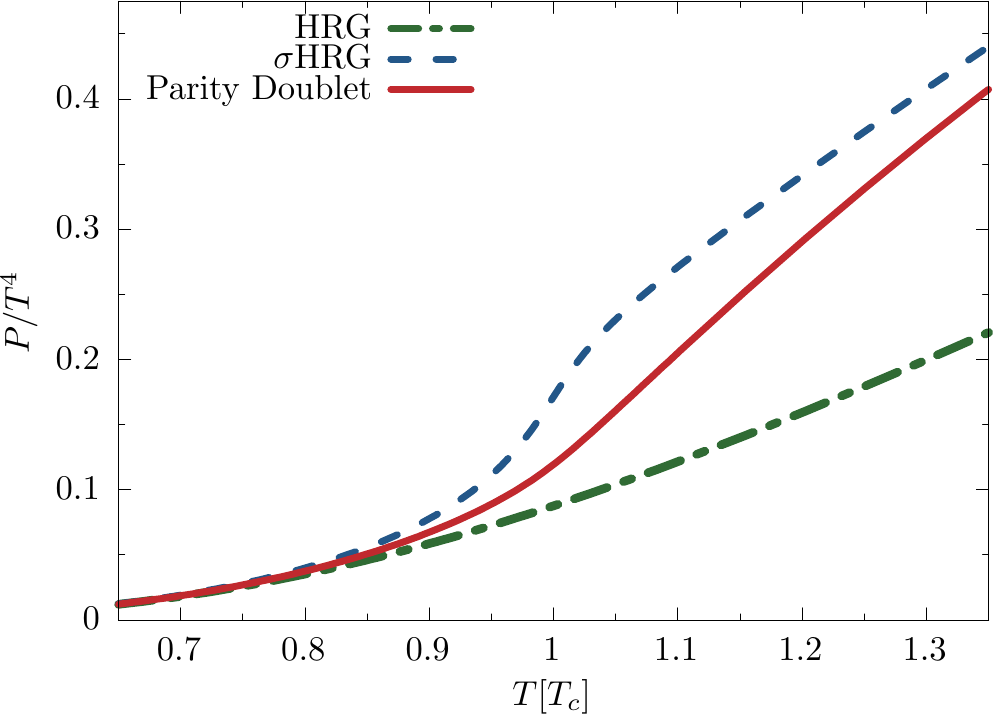}
		\includegraphics[width=0.49\linewidth]{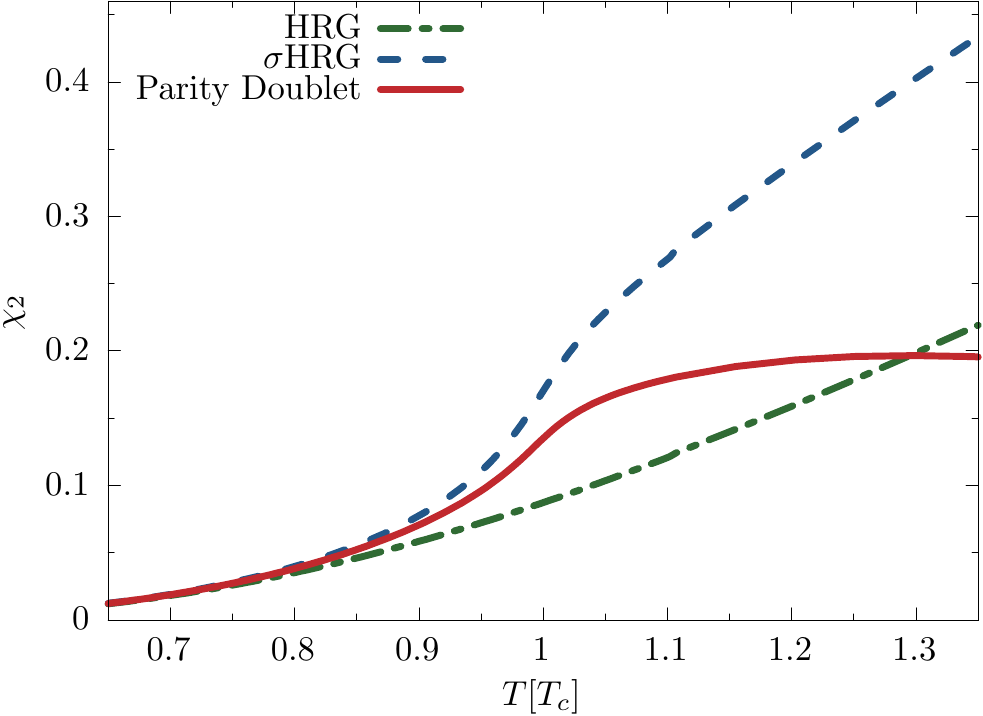}\\
		\caption{Thermodynamic pressure (left), and the second-order cumulant of the net-baryon number density (right), calculated under different schemes, see text. The temperature is expressed in the units of the chiral-critical temperature,  $T_c=T_c(\mu_B=0)$.}
		\label{fig:cumulants}
	\end{figure*}

\section{Higher-order cumulants of the net-baryon number}
\label{sec:cumulants}

	The fluctuations of conserved charges reveal more information about the matter composition than the equation of state and can be used as probes of a phase boundary. The critical properties of chiral models, within the functional renormalization group (FRG) approach~\cite{Wetterich:1992yh, Morris:1993qb, Ellwanger:1993mw, Berges:2000ew}, are governed by the same universality classes as QCD, i.e., the chiral transition belongs to $O(4)$ universality class, which, at large values of the baryon chemical potential, may develop a $Z(2)$ critical point, followed by the first-order phase transition~\cite{Asakawa:1989bq, Halasz:1998qr, Berges:1998rc}. This criticality is naturally encoded in the hadronic parity doublet model, as well as in quark-based models~\cite{Skokov:2010wb, Skokov:2010uh, Schaefer:2006ds, Friman:2011pf, Almasi:2017bhq}, although the mean-field treatment yields different critical exponents.

	The main objective of the present studies is to analyze and delineate the contribution to thermodynamics from chiral dynamics and repulsive baryon-baryon interactions. To this end, we analyze the fluctuations of the net-baryon number at finite temperature and vanishing chemical potential.

	In the grand canonical ensemble, the cumulants of the net-baryon number, $\chi_n$, are commonly defined as temperature-normalized derivatives w.r.t. the baryon chemical potential,
	\begin{equation}\label{eq:fluct_def}
		\chi_n\left(T, \mu_B\right) = T^{n-4} \frac{\partial^{n-1} n_B\left(T, \mu_B\right)}{\partial \mu_B^{n-1}} \textrm,
	\end{equation}
	where $n_B$ is defined in Eq.~\eqref{eq:nb}.

	In the mean-field approximation, the net-baryon number density, as well as any other thermodynamic quantity, contains explicit dependence on the mean fields. Here, we consider only $\sigma$ and $\omega$ mean fields (cf.~Eq.~\eqref{eq:thermo_potential}), thus \mbox{$n_B = n_B\left(T, \mu_B, \sigma(T, \mu_B), \omega(T, \mu_B)\right)$}. 
	Consequently, from Eq.~\eqref{eq:fluct_def} one derives the following general form of the second-order cumulant,
	\begin{equation}\label{eq:x2_all}
		\chi_2 = \chi_2^{\rm id}\beta_{\rm rep} + \frac{\pp n_B}{\pp \sigma}\frac{\pp \sigma}{\pp \mu_B}\textrm,
	\end{equation}
	where $\chi_2^{\rm id} = \chi_2^{\rm id}\left(T, \mu_B, \sigma(T, \mu_B), \omega(T, \mu_B)\right)$ is the ideal gas expression for the net-baryon number susceptibility, and 
	\begin{equation}
		\beta_{\rm rep} = 1 - g_\omega\frac{\partial \omega}{\partial \mu_B}
	\end{equation}
	is the suppression factor due to repulsive interactions. From Eq.~\eqref{eq:x2_all} it is clear that the non-interacting ideal gas result is retrieved when the mean-field contribution is neglected.

	At vanishing chemical potential, Eq.~\eqref{eq:x2_all} reduces to
	\begin{equation}\label{eq:x2_leading}
		\chi_2 = \chi_2^{\rm id}\beta_{\rm rep}\textrm.
	\end{equation}
	We note that, depending on the details of the model, $\chi_2^{\rm id}$ in Eqs.~\eqref{eq:x2_all}~and~\eqref{eq:x2_leading} contains also dependence on the $\sigma$ and $\omega$ mean fields. However, at vanishing $\mu_B$, the expectation value of $\omega$ vanishes as well, i.e., the effective chemical potential is $\mu^\ast = 0$. Thus, $\chi_2^{\rm id}$ contains only the contribution from the $\sigma$ mean field. Therefore, it encodes the information about attractive interactions, while the information about repulsive interactions is contained in the suppression factor $\beta_{\rm rep}$. To some extent, such separation is qualitatively similar to that of the excluded volume approach. We note that the expression for the second-order cumulant in Eq.~\eqref{eq:x2_leading} is exact at vanishing chemical potential.

	In similar spirit, one derives the higher-order cumulants as
	\begin{equation}\label{eq:xn_leading}
	\chi_{n} = \chi_{n}^{\rm id} \beta_{\rm rep}^{n-1}  + \ldots
	\end{equation}
	where $\chi_n^{\rm id}$ is the ideal gas expression for the n'th order cumulant. For $n>2$, Eq.~\eqref{eq:xn_leading} contains extra  terms, as explained in Appendix A. Nevertheless, keeping the first term provides a relatively good approximation to the full expression~\footnote{In Appendix~\ref{sec:appendix}, we present the evaluation of the higher-order cumulants and discuss the comparison of the full expressions and approximations used in this study.}. We note that the general structure of $\chi_n$ derived in Eq.~\eqref{eq:xn_leading} is the same in any kind of $\sigma-\omega$ model under the mean-field approximation and is independent of the details of the model. 

	From Eq.~\eqref{eq:xn_leading}, keeping only the first term, we may also estimate the ratio of the cumulants:
	\begin{equation}\label{eq:xnxm_leading}
	\frac{\chi_n}{\chi_m} = \frac{\chi_{n}^{\rm id}}{\chi_m^{\rm id}} \beta_{\rm rep}^{n-m} + \ldots
	\end{equation}

	Clearly, the separation of the attractive and repulsive contributions persists in the approximation of the higher-order cumulants, as well as in their ratios. This allows to precisely delineate the contribution of chiral symmetry restoration and repulsive interaction to the critical behavior of the cumulants in the vicinity of the chiral phase transition.
	
	In the following, we quantify and discuss the properties of the obtained equation of state, as well as the higher-order cumulants of the net-baryon number density at vanishing chemical potential in order to identify the importance of the repulsive interactions near the chiral crossover transition. 

\section{Results}
\label{sec:results}

	In the left panel of Fig.~\ref{fig:cumulants}, we show the numerical results on the temperature-normalized thermodynamic pressure at vanishing baryon chemical potential. The pressure obtained in the HRG model increases monotonically and does not resemble any critical behavior. This is expected because  $\Omega^{\rm HRG}$ is just a sum of uncorrelated particles (cf. Eq.~\eqref{eq:hrg_thermo}) with vacuum hadron masses. There are clear deviations of the parity doublet model result on thermodynamic pressure from the corresponding ideal HRG. The increase of the pressure is a bulk consequence of an interplay between critical chiral dynamics with in-medium hadron masses and repulsive interactions. The influence of in-medium hadron masses is identified when considering the pressure $P^{\sigma \rm HRG}$ of the $\sigma$HRG model which increases around $T_c$. This is evidently linked to the in-medium shift of baryon masses due to chiral symmetry restoration. We note,  that  $P^{\sigma \rm HRG}$ is systematically higher than the parity doublet pressure. The reason is that the partition function of uncorrelated particles does not contain the mean-field potentials which provide a negative contribution to the pressure,  as seen in the parity doublet model from Eq.~\eqref{eq:thermo_potential}. All pressures shown in Fig.~\ref{fig:cumulants}  converge to the Stefan-Boltzmann limit at high temperatures.

    In the right panel of Fig.~\ref{fig:cumulants}, we show the second-order cumulant, $\chi_2$. We note that, at vanishing chemical potential the expectation value of $\omega$ is zero, thus  $\chi_n^{\rm id}$ are equivalent to the $\sigma$HRG formulation. Similarly to the case of pressure, the result for $\sigma$HRG lies systematically above the ideal gas result. $\chi_2$ in  HRG and $\sigma$HRG  models converge to the  Stefan-Boltzmann limit at high-temperatures. In contrast, the parity doublet result saturates above $T_c$ and monotonically decreases to zero at high temperature. From Eq.~\eqref{eq:x2_leading}, it is clear that the difference between $\sigma$HRG and parity doublet results is due to the suppression originating from $\beta_{\rm rep}$.

\begin{figure}[t!]
		\centering
		\includegraphics[width=\columnwidth]{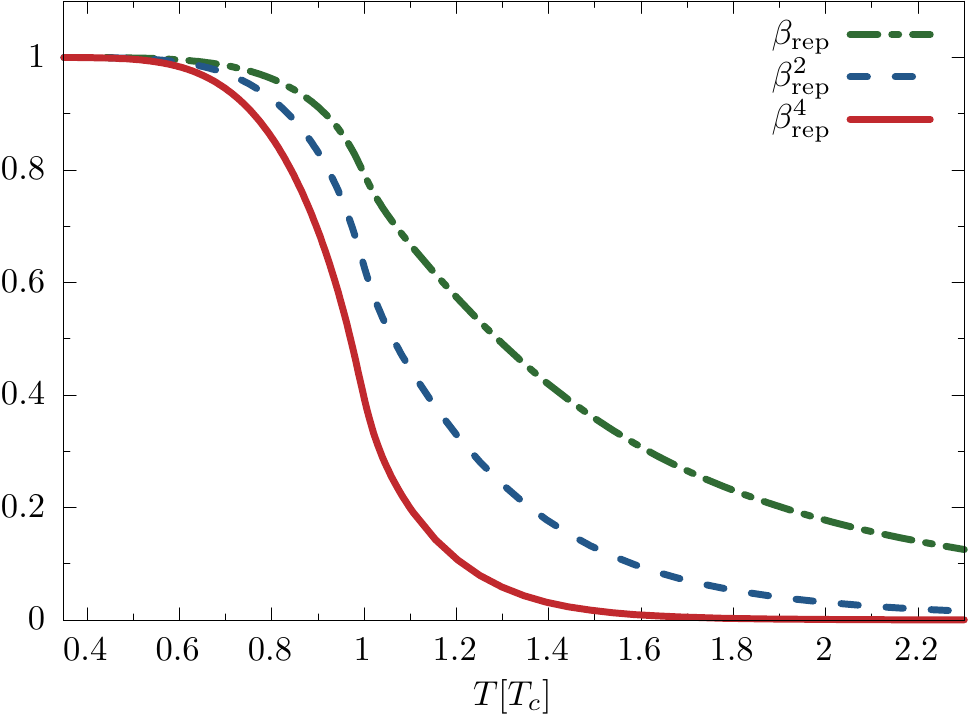}
		\caption{The suppression factor $\beta_{\rm rep}$ from Eq.~\eqref{eq:xn_leading}  for different temperatures, expressed in the units of the chiral-critical temperature,  $T_c=T_c(\mu_B=0)$.}
		\label{fig:suppression}
	\end{figure}

	In Fig.~\ref{fig:suppression}, we show the suppression factor, $\beta_{\rm rep}$. It changes gradually from unity at low temperatures to zero at high temperatures. This indicates that the repulsive forces become more important with increasing temperature. At the critical temperature, $\beta_{\rm rep}\simeq 0.8$. Consequently, $\chi_2^{\rm id}$ is reduced by $20\%$ due to repulsive interactions. From Eq.~\eqref{eq:xn_leading} one may also estimate the suppression of the higher-order cumulants. At $T\simeq T_c$, the suppression of $\chi_4$ and $\chi_6$ due to repulsive interactions amounts to $41\%$ and $67\%$, respectively.
	
	\begin{figure*}[t!]
		\centering
		\includegraphics[width=0.49\linewidth]{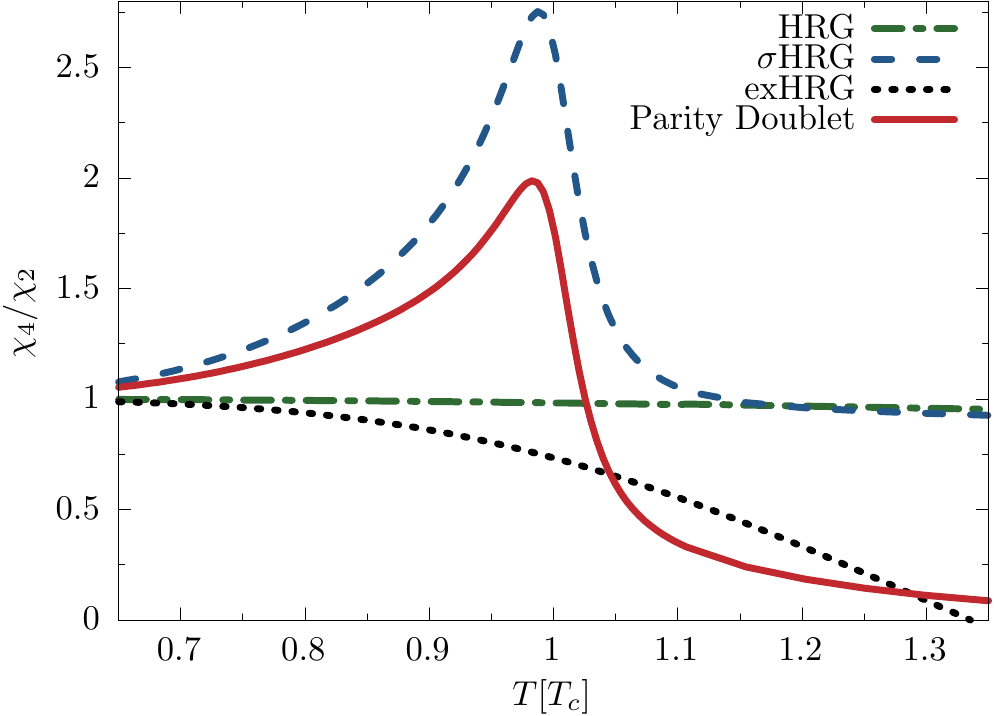}
		\includegraphics[width=0.49\linewidth]{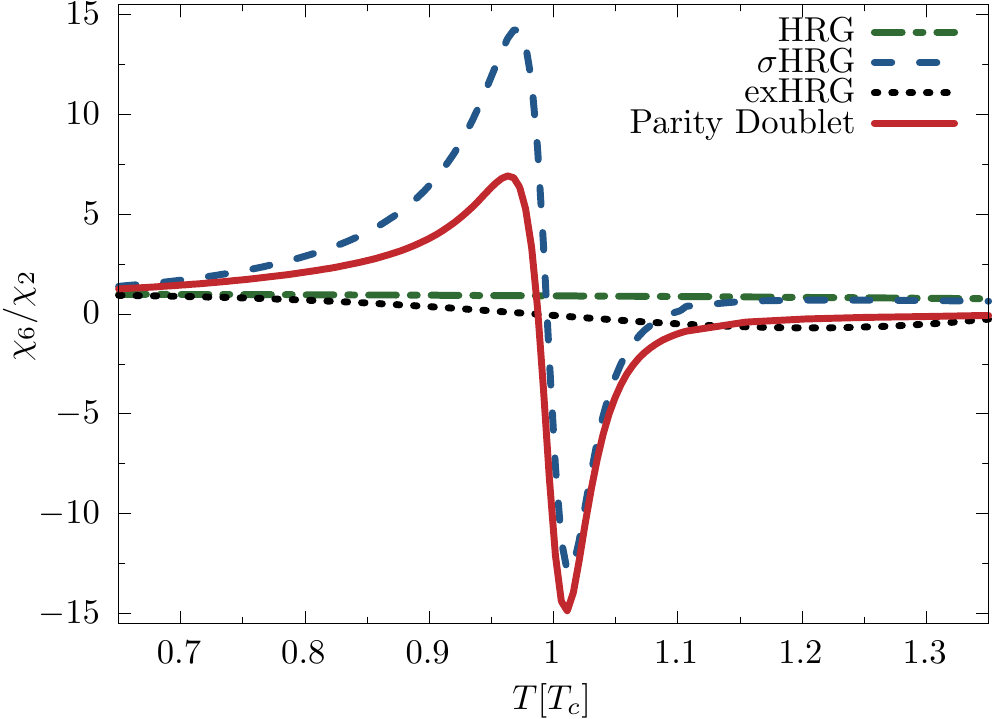}
		\caption{Ratio of different  higher-order cumulants of the net-baryon number density fluctuations, $\chi_4/\chi_2$ (left) and $\chi_6/\chi_2$ (right),
		calculated under different schemes, see text. The temperature is expressed in the units of the chiral-critical temperature, $T_c=T_c(\mu_B=0)$.
		}
		\label{fig:ratios}
	\end{figure*}

	In the left panel of Fig.~\ref{fig:ratios}, we show the net-baryon kurtosis $\chi_4/\chi_2$, and the ratio $\chi_6/\chi_2$. For the ideal HRG model, these ratios are equal to unity due to the expected Skellam probability distribution of the net-baryon density~\cite{BraunMunzinger:2003zd}. The chiral dynamics and repulsive interactions implemented in the parity doublet model imply strong deviations of these fluctuation ratios from the Skellam baseline. The kurtosis exhibits a peak around the transition temperature, after which it drastically drops below unity. This is in contrast to the $\sigma$HRG result, where the peak structure appears as well,  however, the result converges back to the Skellam distribution limit at higher temperatures. Thus, the appearance of the peak in the kurtosis is attributed to remnants of the chiral symmetry restoration, whereas strong suppression around $T\simeq T_c$ is due to repulsive interactions between baryons. We note that at vanishing chemical potential and in the chiral limit $\chi^n$ cumulants are non-critical up to $n<6$ order~\cite{Friman:2011pf}. Thus, the peak of the kurtosis stays finite at the chiral phase transition and its strength is strongly model-dependent. It is known that when including quantum and thermal fluctuations in effective chiral models within the FRG approach such a peak essentially disappears leaving the kurtosis nearly unaffected by the chiral symmetry restoration~\cite{Skokov:2010wb, Skokov:2010uh, Schaefer:2006ds, Friman:2011pf}. The influence of repulsive interactions, however, can imply suppression of kurtosis near the chiral crossover below the Skellam baseline. In this model calculations, such suppression at $T\simeq T_c$ is of the order of $35\%$. The kurtosis of net baryon-number density was introduced as an excellent probe of quark deconfinement, and outside the critical region was shown to be quantified by the square of the baryon number carried by medium constituents~\cite{Ejiri:2005wq, Karsch:2005ps}. Such an interpretation of kurtosis is not accessible in the present hadronic chiral model since it does not contain quarks degrees of freedom. Thus, the parity doublet model can be only applicable up to the near vicinity of the chiral crossover.

	The ratio $\chi_6/\chi_2$ exhibits a strong sensitivity to dynamical effects related to chiral symmetry restoration, as shown in the right panel of Fig.~\ref{fig:ratios}. The characteristic S-type structure of this ratio obtained in the parity doublet model with a well-pronounced peak followed by a dip at negative values in the near vicinity of $T_c$ is expected as an imprint of the chiral criticality ~\cite{Friman:2011pf}. The leading role of the chiral symmetry restoration on the properties of $\chi_6/\chi_2$   is also seen by comparing the full parity doublet and $\sigma$HRG  model results in Fig.~\ref{fig:ratios}. In both cases, the S-type structure of this ratio is preserved, albeit with some quantitative differences which are linked to repulsive interactions. Indeed, as already discussed in the context of the kurtosis, the presence of repulsive interactions suppresses $\chi_6/\chi_2$ when compared to the $\sigma$HRG results. Nevertheless, the qualitative structure of this ratio remains the same. 

    The magnitude of the peak-dip structure observed in $\chi_6/\chi_2$ is a direct consequence of the mean-field approximation employed in our calculations. We note that the inclusion of mesonic fluctuations weakens the critical behavior. This was presented in other models exhibiting $O(4)$ chiral criticality, e.g., quark-meson (QM)~\cite{Morita:2014fda} and Polyakov loop-extended quark-meson (PQM)~\cite{Skokov:2010wb, Skokov:2010uh, Friman:2011pf, Almasi:2017bhq} models within FRG approach. In these models, the $\chi_4/\chi_2$  ratio decreases monotonously with temperature and practically no peak structure is exhibited neat $T_c$. In contrast, the general peak-dip structure of $\chi_6/\chi_2$ obtained in the mean-field approximation prevails when mesonic fluctuations are included. This highly non-monotonic behavior is also seen in lattice QCD simulations. In particular, the peak in $\chi_6/\chi_2$ is seen, despite huge systematic error at low temperatures~\cite{Bazavov:2017dus, Borsanyi:2018grb}.

	Lastly, we compare the properties of $\chi_4/\chi_2$  and $\chi_6/\chi_2$ fluctuation ratios with the excluded volume formulation of the repulsive interactions. The effect of excluded volume on thermodynamic properties of a hadronic medium was extensively studied~\cite{Andronic:2012ut, Rischke:1991ke, Albright:2014gva, Albright:2015uua, Zalewski:2015yea, Lo:2017ldt, Venugopalan:1992hy}. We consider the common formulation of the excluded volume effect, in which it is considered for the bulk pressure of the system. In Ref.~\cite{Lo:2017ldt}, it was pointed out that this may not be a robust gauge of the repulsion in the individual interaction channels.
	
	The thermodynamic pressure in the excluded volume approach is given by a thermodynamically self-consistent equation:
	\begin{equation}\label{eq:press_exvol}
		P^{\rm ex}(T, \mu^\star) = \sum_k P^{\rm id}\left(T, \mu_k\right) \textrm,
	\end{equation}
	with $\mu_k = \pm\mu_B - v_0 P^{\rm ex}(T, \mu^\star)$, and  $P^{\rm id}$ is the ideal-gas pressure. The ($\pm$) sign applies to particles and antiparticles, respectively. The summation over $k$ goes over particles and anti-particles. The constant eigenvolume, 
	\begin{equation}
		v_0 = \frac{16}{3} \pi r_0^3\textrm, 
	\end{equation}
	where $r_0 = 0.3~$fm. Once the excluded volume, $v_0$, is specified, the pressure can be obtained by solving Eq.~\eqref{eq:press_exvol} self-consistently. From this, higher-order cumulants are obtained numerically.

	In Fig.~\ref{fig:ratios}, we show the  ratios of cumulants,  $\chi_4/\chi_2$ and $\chi_6/\chi_2$,  obtained in the excluded volume approach labeled as exHRG. For consistency, the temperature is normalized to the critical temperature, $T_c$, obtained in the parity doublet model. The excluded volume model shows a swift decrease from the ideal HRG behavior at low temperature and turns negative at around $1.35~T_c$. Recently, it was suggested that this behavior may call into question the connection to deconfinement transition in QCD~\cite{Vovchenko:2016rkn}. For $\chi_6/\chi_2$ the excluded volume approach deviates from the ideal HRG result, turns negative, and predicts a dip above $T_c$. Very similar behavior is also observed in models incorporating Van-der-Waals type formulation of repulsive and attractive interactions between hadrons~\cite{Vovchenko:2016rkn}.
	
	Clearly, the results of the excluded volume and the parity doublet model are qualitatively different. However, our consistent chiral model calculations confirmed that indeed repulsive interactions between hadrons imply suppression of $\chi_4/\chi_2$ and $\chi_6/\chi_2$ fluctuation ratios near the chiral crossover as pointed out in Ref.~\cite{Vovchenko:2016rkn}. Thus, the phenomenological hadronic models that account for repulsive and attractive interactions between constituents can be successful in describing some deviations of net-charge fluctuations from the HRG baseline observed in LQCD. However, this is not the case if such fluctuation observables are affected by the chiral criticality. This is very transparent when considering $\chi_6/\chi_2$ ratio shown in Fig.~\ref{fig:ratios}. The lack of in-medium effects due to the chiral symmetry restoration in the excluded volume approach directly implies that such a model is not capable of reproducing a characteristic structure of this ratio near the chiral crossover. This indicates that in order to fully describe the properties of cumulants of net-baryon number fluctuations near the chiral crossover, it is not sufficient to account only for repulsive interactions, but it is essential to formulate a consistent framework that implements the chiral in-medium effects and repulsive interactions simultaneously.

\section{Summary}
\label{sec:summary}

	We have discussed the role of attractive and repulsive nucleon-nucleon interactions on the thermodynamic and chiral-critical properties of a strongly interacting medium at finite temperature. To this end, we have used the parity doublet model in the mean-field approximation. We have analyzed the thermodynamic pressure and the cumulants of the net-baryon number density, $\chi_n$, up to the sixth order, as well as their ratios.

	We have shown that, at vanishing chemical potential, the second-order cumulant factorizes as a product of a term that is directly linked to attractive scalar interactions and a suppression factor due to repulsive interactions. Furthermore, we have found that to a good approximation, a  similar separation also holds for higher-order cumulants. This allowed to consistently delineate different in-medium effects and to identify the role of repulsive interactions near the chiral crossover transition. 
	
	We have pointed out that even a moderate influence of the repulsive interactions between hadrons on the equation of state becomes more readily exposed in the quantitative structure of the cumulants with increasing their order. Consequently, in the phenomenological description of cumulants of net baryon density fluctuations calculated in LQCD or measured in heavy-ion collisions attention is to be given to account for possible repulsive interactions between baryons. 
	
	A frequently used approach to account for repulsive hadronic interactions is the hard-core repulsion or van-der-Waals type interaction model. We have compared our results for the n$^{\rm th}$-order cumulants of net-baryon number fluctuations with an excluded volume formulation of the repulsive interactions. Such phenomenological model provides suppression of cumulants with increasing temperature due to hadronic repulsion. In particular, the kurtosis  $\chi_4/\chi_2$ in this model is reduced from the Skellam limit towards the chiral crossover, as observed in LQCD results. However, when considering the $\chi_6/\chi_2$ fluctuation ratio, which exhibits a dominant contribution from the chiral criticality, such phenomenological model fails to capture the characteristic properties of this ratio. Consequently, a description of the higher-order cumulants of the net-baryon density in the chiral crossover requires a consistent framework that accounts for a self-consistent treatment of the chiral in-medium effects and repulsive interactions.

    At low temperature, the parity doublet model predicts sequential liquid-gas and chiral phase transitions in the baryon-rich matter, with critical endpoints of both transitions at moderate temperatures of tens of MeV~\cite{Zschiesche:2006zj, Sasaki:2010bp}. It is challenging to identify the role of in-medium effects and hadronic interactions on the properties of higher-order cumulants near these distinct phase transitions. Work in this direction is in progress.

\begin{acknowledgements}
       This work was partly supported by the Polish National Science Center (NCN), under OPUS Grant No. 2018/31/B/ST2/01663 (K.R. and C.S.) and Preludium Grant No. UMO-2017/27/N/ST2/01973 (M.M.). K.R. also acknowledges the support of the Polish Ministry of Science and Higher Education.  M.M. acknowledges helpful comments from Pok Man Lo.
\end{acknowledgements}

\appendix

\section{Evaluation of the higher-order cumulants}
\label{sec:appendix}

\begin{figure*}[t!]
	\centering
	\includegraphics[width=\columnwidth]{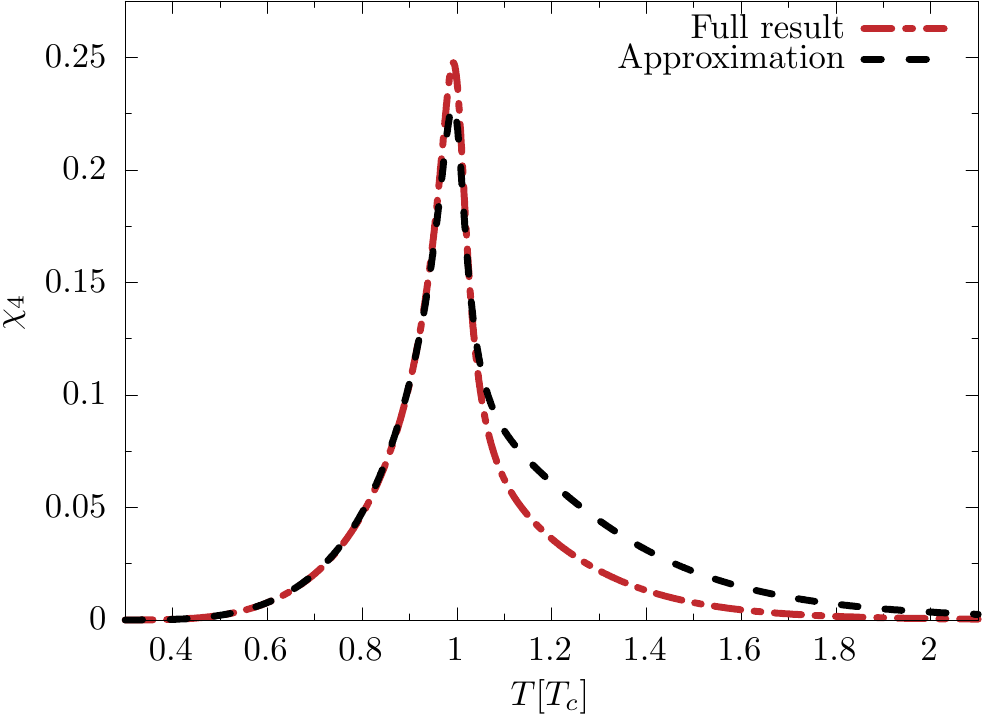}
	\includegraphics[width=\columnwidth]{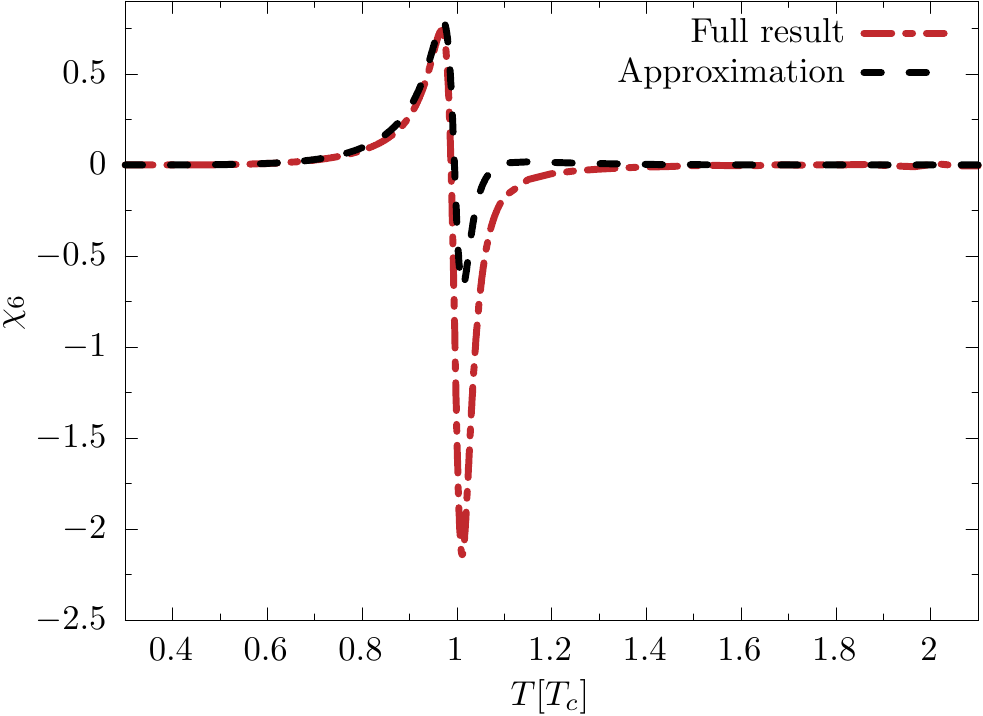}\\
	\caption{Fourth- and sixth-order cumulants of the net-baryon number obtained in the parity doublet model and their approximations at $\mu_B=0$, see text. The temperature is expressed in the units of the chiral-critical temperature, $T_c=T_c(\mu_B=0)$.}
	\label{fig:approx_cumulants}
\end{figure*}

From the definition of the n$^{\rm th}$ order cumulant of the net-baryon number density,

\begin{equation}
\chi_{n} = T^{n-4}\frac{\pp^{n-1} n_B}{\pp \mu_B^{n-1}} \textrm, 
\end{equation}
one can derive the following relation:
\begin{widetext}
\begin{equation}\label{eq:xn_trick}
\begin{split}
\chi_{n+1} &=\frac{\pp \chi_n \left(T, \mu_B, \sigma(T,\mu_B), \omega(T,\mu_B)\right) }{\pp \mu_B} = \frac{\pp \chi_n}{\pp \mu_B} + \frac{\pp \chi_n}{\pp \omega}\frac{\pp \omega}{\pp \mu_B} + \frac{\pp \chi_n}{\pp \sigma}\frac{\pp \sigma}{\pp \mu_B} =\\ &\frac{\pp \chi_n}{\pp \mu^\ast}\frac{\pp \mu^\ast}{\pp \mu_B} + \frac{\pp \chi_n}{\pp \mu^\ast}\frac{\pp \mu^\ast}{\pp \omega}\frac{\pp \omega}{\pp \mu_B} + \frac{\pp \chi_n}{\pp \sigma}\frac{\pp \sigma}{\pp \mu_B}
= \chi_{n+1}^{\rm id} - g_\omega\chi_{n+1}^{\rm id} \frac{\pp \omega} {\pp \mu_B} + \frac{\pp\chi_n}{\pp \sigma} \frac{\pp \sigma}{\pp \mu_B} = \chi_{n+1}^{\rm id}\beta_{\rm rep} + \frac{\pp\chi_n}{\pp \sigma} \frac{\pp \sigma}{\pp \mu_B} \textrm,
\end{split}
\end{equation}
\end{widetext}
where $\chi_{n+1}^{\rm id}$ is the ideal gas expression for the (n+1)$^{\rm th}$ order cumulant, $\beta_{\rm rep} =  1 - g_\omega \pp \omega / \pp \mu_B$, and $\mu^\ast = \mu_B - g_\omega \omega$ is the effective chemical potential. Applying Eq.~\eqref{eq:xn_trick} to the net-baryon density, one obtains the second-order cumulant as
\begin{equation}\label{eq:x2_app}
\chi_2 = \chi_2^{\rm id}\beta_{\rm rep} + \frac{\pp n_B}{\pp \sigma}\frac{\pp \sigma}{\pp \mu_B} \textrm.
\end{equation}
At $\mu_B=0$, $n_B$ vanishes, owing to particle-antiparticle symmetry. Hence, the last term in Eq.~\eqref{eq:x2_app} vanishes and one obtains Eq.~\eqref{eq:x2_leading}. 

Higher-order cumulants are obtained by applying Eq.~\eqref{eq:xn_trick} iteratively. In particular, the third- and fourth-order cumulants, read 
\begin{widetext}
\begin{subequations}\label{eq:x3_app2}
\begin{align}
\chi_3 =& \chi_3^{\rm id}\beta^2_{\rm rep} + 2 \frac{\pp \chi_2^{\rm id}}{\pp \sigma}\frac{\pp \sigma}{\pp \mu_B}\beta_{\rm rep} -g_\omega\chi_2^{\rm id}\frac{\pp^2 \omega}{\pp \mu_B^2} + \frac{\pp^2 n_B}{\pp \sigma^2}\left(\frac{\pp \sigma}{\pp \mu_B}\right)^2 + \frac{\pp n_B}{\pp \sigma}\frac{\pp^2\sigma}{\pp\mu_B^2} \textrm, \\
\chi_4 =& \chi_4^{\rm id}\beta_{\rm rep}^3 + 3\frac{\pp \chi_3^{\rm id}}{\pp \sigma}\frac{\pp \sigma}{\pp \mu_B}\beta_{\rm rep}^2 - 3g_\omega \chi_3^{\rm id}\frac{\pp^2 \omega}{\pp\mu_B^2}\beta_{\rm rep} + 3\frac{\pp^2 \chi_2^{\rm id}}{\pp \sigma^2}\left(\frac{\pp \sigma}{\pp \mu_B}\right)^2\beta_{\rm rep}\nonumber\\
 &+ 3\frac{\pp \chi_2^{\rm id}}{\pp \sigma}\left(\frac{\pp^2\sigma}{\pp\mu_B^2}\beta_{\rm rep} - g_\omega \frac{\pp\sigma}{\pp\mu_B}\frac{\pp^2\omega}{\pp\mu_B^2}\right) - g_\omega\chi_2^{\rm id} \frac{\pp^3\omega}{\pp\mu_B^3} + \frac{\pp^3 n_B}{\pp\sigma^3}\left(\frac{\pp \sigma}{\pp \mu_B}\right)^3    + 3\frac{\pp^2 n_B}{\pp \sigma^2}\frac{\pp\sigma}{\pp \mu_B}\frac{\pp^2 \sigma}{\pp\mu_B^2} + \frac{\pp n_B}{\pp \sigma}\frac{\pp^3\sigma}{\pp \mu_B^3}.
\end{align}
\end{subequations}
\end{widetext}

We note,  that $\sigma(T,\mu_B) = \sigma(T,-\mu_B)$ and $\omega(T,\mu_B) = - \omega(T,-\mu_B)$. {Thus,}  the odd derivatives of $\sigma$ and even derivatives of $\omega$ w.r.t $\mu_B$ vanish, i.e.,
\begin{equation}
\frac{\pp^{2k+1} \sigma}{\pp \mu_B^{2k+1}}\Bigg|_{\mu_B=0} = 0 \textrm,
~~~~~\frac{\pp^{2k} \omega}{\pp \mu_B^{2k}}\Bigg|_{\mu_B=0} = 0 \textrm.
\end{equation}

Consequently, at $\mu_B=0$, each term in $\chi_3$ vanishes separately, thus $\chi_3(T,\mu_B=0) = 0$.  On the other hand, the even-order cumulants, $\chi_2$ and $\chi_4$, are reduced to
\begin{subequations}
\begin{align}
\chi_2 &= \chi_2^{\rm id}\beta_{\rm rep} \textrm, \\
\chi_4 &= \chi_4^{\rm id}\beta_{\rm rep}^3 + 3\frac{\pp \chi_2^{\rm id}}{\pp \sigma} \frac{\pp^2\sigma}{\pp \mu_B^2}\beta_{\rm rep} - g_\omega \chi_2^{\rm id}\frac{\pp^3\omega}{\pp\mu_B^3} \textrm.\label{eq:x4_app}
\end{align}
\end{subequations}

The higher-order cumulants are calculated following a similar method as explained above. The approximations used in Eqs. \eqref{eq:x2_leading} and \eqref{eq:xn_leading} are obtained by dropping all terms except the first one.  In Fig.~\ref{fig:approx_cumulants}, we present a direct comparison of the exact results for the fourth- and sixth-order cumulants and their approximations given in Eq.~\eqref{eq:xn_leading}. The agreement is quite satisfactory for temperatures $T\leq 1.1~T_c$, keeping the basic properties of the cumulants.

\bibliography{biblio}

\end{document}